\documentclass[10pt,twocolumn]{article}

\usepackage[utf8]{inputenc}
\usepackage{amsmath}
\usepackage{amssymb}
\usepackage{amsthm}
\usepackage{titlesec}
\usepackage{cite}
\usepackage{color}
\usepackage{booktabs}
\usepackage{graphicx}
\usepackage{nicefrac}
\usepackage[colorlinks=false]{hyperref}
\usepackage[format=plain,labelfont=it]{caption}
\usepackage[left=1.5cm,right=1.5cm,top=2cm,bottom=2cm]{geometry}
\usepackage{algpseudocode}
\usepackage[FIGTOPCAP]{subfigure}

\titleformat{\section}{\centering\normalfont\scshape}{\Roman{section}.}{5pt}{}
\titleformat{\subsection}{\normalfont\it}{\Alph{subsection}.}{5pt}{}
\titleformat{\subsubsection}{\normalfont\it}{\hspace{4mm}\arabic{subsubsection})}{5pt}{}

\newcommand\infoFootnote[1]{%
  \begingroup
  \renewcommand\thefootnote{}\footnote{#1}%
  \addtocounter{footnote}{-1}%
  \endgroup}

\newtheorem{thm}{Theorem}
\newtheorem{cor}[thm]{Corollary}
\newtheorem{lem}[thm]{Lemma}
\newtheorem{prop}[thm]{Proposition}

\newtheorem{rem}{Remark}
\newtheorem{alg}{Algorithm}

\newcommand{\R}{\mathbb{R}} 
\newcommand{\N}{\mathbb{N}} 

\newcommand{\Ac}{\mathcal{A}}
\newcommand{\Bc}{\mathcal{B}}
\newcommand{\Fc}{\mathcal{F}}
\newcommand{\Pc}{\mathcal{P}}
\newcommand{\Rc}{\mathcal{R}}

\newcommand{\Ic}{\mathcal{I}}
\newcommand{\Tc}{\mathcal{T}} 
\newcommand{\Uc}{\mathcal{U}} 
\newcommand{\Xc}{\mathcal{X}} 

\newcommand{\Ab}{\boldsymbol{A}}
\newcommand{\Bb}{\boldsymbol{B}}
\newcommand{\Cb}{\boldsymbol{C}}

\newcommand{\Hb}{\boldsymbol{H}}
\newcommand{\Ib}{\boldsymbol{I}}

\newcommand{\Pb}{\boldsymbol{P}}
\newcommand{\Qb}{\boldsymbol{Q}}
\newcommand{\Rb}{\boldsymbol{R}}
\newcommand{\Sb}{\boldsymbol{S}}

\newcommand{\Ub}{\boldsymbol{U}}

\newcommand{\Wb}{\boldsymbol{W}}
\newcommand{\Xb}{\boldsymbol{X}}

\newcommand{\bb}{\boldsymbol{b}}
\newcommand{\cb}{\boldsymbol{c}}

\newcommand{\fb}{\boldsymbol{f}}
\newcommand{\gb}{\boldsymbol{g}}
\newcommand{\hb}{\boldsymbol{h}}

\newcommand{\pb}{\boldsymbol{p}}
\newcommand{\qb}{\boldsymbol{q}}

\newcommand{\ub}{\boldsymbol{u}}
\newcommand{\vb}{\boldsymbol{v}}

\newcommand{\xb}{\boldsymbol{x}}
\newcommand{\yb}{\boldsymbol{y}}
\newcommand{\zb}{\boldsymbol{z}}

\newcommand{\xib}{\boldsymbol{\xi}}
\newcommand{\pib}{\boldsymbol{\pi}}
\newcommand{\Phib}{\boldsymbol{\Phi}}

\newcommand{\gammab}{\boldsymbol{\gamma}}

\newcommand{\deltab}{\boldsymbol{\delta}}
\newcommand{\kappab}{\boldsymbol{\kappa}}

\newcommand{\zerob}{\boldsymbol{0}}
\newcommand{\oneb}{\boldsymbol{1}}

\newcommand{\interior}{\mathrm{int}}

\newcommand{\norm}[1]{\left\lVert#1\right\rVert}

\title{\vspace{-2mm}\bf Reachability analysis for piecewise affine systems with\\ neural network-based controllers$^\ast$}
\author{Dieter Teichrib and Moritz Schulze Darup\vspace{2mm}}
\date{}

\begin{document}

\maketitle

\textbf{\textit{Abstract}.} {\bf Neural networks (NN) have been successfully applied to approximate various types of complex control laws, resulting in low-complexity NN-based controllers that are fast to evaluate. However, when approximating control laws using NN, performance and stability guarantees of the original controller may not be preserved. Recently, it has been shown that it is possible to provide such guarantees for linear systems with NN-based controllers by analyzing the approximation error with respect to a stabilizing base-line controller or by computing reachable sets of the closed-loop system. The latter has the advantage of not requiring a base-line controller. In this paper, we show that similar ideas can be used to analyze the closed-loop behavior of piecewise affine (PWA) systems with an NN-based controller. Our approach builds on computing over-approximations of reachable sets using mixed-integer linear programming, which allows to certify that the closed-loop system converges to a small set containing the origin while satisfying input and state constraints. We also show how to modify a given NN-based controller to ensure asymptotic stability for the controlled PWA system.}
\infoFootnote{D. Teichrib and M. Schulze Darup are with the \href{https://rcs.mb.tu-dortmund.de/}{Control and~Cyber-physical Systems Group}, Faculty of Mechanical Engineering, TU Dortmund University, Germany. E-mails:  \href{mailto:moritz.schulzedarup@tu-dortmund.de}{\{dieter.teichrib, moritz.schulzedarup\}@tu-dortmund.de}. \vspace{0.5mm}}
\infoFootnote{\hspace{-1.5mm}$^\ast$This paper is a \textbf{preprint} of a contribution to the 63rd IEEE Conference on Decision and Control 2024.}

\section{Introduction}
Piecewise affine (PWA) systems have emerged as an important system class in the control 
literature as they can be used to model or approximate a large class of hybrid systems or nonlinear systems \cite{Azuma2010}. Moreover, the equivalence of PWA systems and mixed logical dynamical (MLD) systems has been shown in \cite{Bemporad2000MLDandPWA}, which allows the use of techniques developed for MLD systems also for PWA systems and vice versa. Stability analysis and control synthesis for PWA systems are an important branch of research, and many methods based on linear matrix inequalities (LMI) have been proposed~{\cite{Mignone2000,Ferrari2001,Ferrari2002,Grieder2005}}. However, these methods are often only applicable to linear segments of the PWA system dynamics that contain the origin. Furthermore, it is not possible to handle constraints with these methods. Therefore, model predictive control (MPC) for PWA systems (see, e.g., \cite{Camacho2010} for an overview) gained attention 
due to the possibility of directly considering constraints in the optimal control problem (OCP). When using MPC for PWA systems, the switching nature of the system dynamics is typically described by mixed-integer linear constraints \cite{Bemporad1999}. Thus, the OCP with a quadratic cost function becomes a mixed-integer quadratic program (MIQP) that must be solved in every time step. This can be challenging for long prediction horizons or PWA systems with many affine segments because the MIQP may be too complex to be solved within the sampling period. As a consequence, methods have been developed to approximate the optimal control law of MPC (\cite{Jones2009,Chen2018}) by functions that are fast to evaluate. A popular choice are neural networks (NN) \cite{Xiang2018}, as they can approximate a large class of functions with arbitrary accuracy. Moreover, NN with PWA activations such as maxout NN \cite{Goodfellow2013} share the PWA structure of the MPC law for PWA systems \cite[Thm.~1]{Borrelli2003}. Unfortunately, stability and constraint satisfaction are typically not guaranteed for the NN-based controller. In addition, NN are known to react unexpectedly even for small perturbations of the input \cite{Szegedy2014}. Still, for linear systems, some methods to certify stability and constraint satisfaction of the NN-based controller exist. The most commonly used ones are based on mixed-integer linear programming (MILP) (\cite{Karg2020Reach,Schwan2023}), semi-definite programming \cite{Fazlyab2022}, or satisfiability modulo theory \cite{Huang2016}. 
However, these methods are either not applicable to PWA systems or require a stabilizing base-line controller. 

In this paper, we propose a method for certifying stability and constraint satisfaction for PWA systems with NN-based controllers without using a base-line controller. Our method is based on the computation of over-approximations of $k$-step reachable sets of the closed-loop system using MILP. Similar ideas have been used in \cite{Karg2020Reach} for linear systems and in \cite{Xiang2018} for safety verification of PWA systems. However, both existing methods are not suitable for certifying stability and constraint satisfaction for PWA systems with NN-based controllers. In contrast, our method allows us to certify that the closed-loop system converges to a small positively invariant (PI) set containing the origin while satisfying input and state constraints. We also show how to modify the NN-based controller to ensure asymptotic stability for the controlled PWA system.

The paper is organized as follows. In the remainder of this section, we state relevant notation and basic definitions. In Section~\ref{sec:Fundamentals}, some fundamentals on MPC for PWA systems and NN with PWA activation are given. Section~\ref{sec:CLAnalysis} contains the main contributions of this paper. Namely, the analysis of the closed-loop system consisting of a PWA system and an NN-based controller by MILP. This MILP is then used to compute PI sets, and over-approximations of reachable sets. 
These sets are the primary tools used to show the stability of the closed-loop system. A case study is given in Section~\ref{sec:Example} to illustrate the rather technical results. Finally, conclusions are presented in Section~\ref{sec:Conclusions}.

\subsection{Notation and definitions}
We define the support function of a polyhedron $\Xc$ for a row-vector $\vb\in\R^{1\times n}$ as $h_{\Xc}(\vb):=\sup_{\xb\in\Xc} \vb\xb$. For a matrix argument $\Hb\in\R^{w\times n}$, $h_{\Xc}(\Hb)$ is understood as $h_{\Xc}(\Hb)=\begin{pmatrix} h_{\Xc}(\Hb_1) & \dots & h_{\Xc}(\Hb_w) \end{pmatrix}^\top$, where $\Hb_i$ refers to the $i$-th row of $\Hb$. Support functions have many useful features. For instance, consider a polyhedron of the form $\Bc:=\{\xb\in\R^n \ | \ \Hb^{(\Bc)} \xb \leq \hb^{(\Bc)} \}$, then $\Ac\subseteq\Bc$ if and only if $h_{\Ac}(\Hb^{(\Bc)})\leq \hb^{(\Bc)}$. The scaling of a set $\Ac$ by a scalar $s$ is defined as $s \Ac:=\{s \xb \in \R^n \ | \ \xb \in \Ac \}$. Natural numbers and natural numbers including $0$ are denoted by $\N$ and $\N_0$, respectively. We further define uniform ultimate boundedness (UUB) of a system with respect to a C-set (i.e., a convex and compact set containing the origin in its interior) according to \cite[Def.~2.4]{Blanchini1994}: A system is denoted as ultimately bounded in a C-set $\Tc$, uniformly in $\Fc$, if for every initial condition $\xb(0)\in\Fc$, there exits a $k^\ast(\xb(0))$ such that $\xb(k)\in\Tc$ holds for all $k\geq k^\ast(\xb(0))$.

\section{Fundamentals of MPC for PWA systems and neural networks with PWA activation}\label{sec:Fundamentals}
Consider a PWA system of the form 
\begin{align}
    \nonumber
    \xb(&k+1)=\fb_{\text{PWA}}(\xb(k),\ub(k)) \\
    \label{eq:PWA_System}
    &:=\Ab^{(i)}\xb(k)\!+\!\Bb^{(i)}\ub(k)\!+\!\pb^{(i)} \,\,\,\text{if} \ \begin{pmatrix}
        \xb(k) \\
        \ub(k)
    \end{pmatrix} \in \Pc^{(i)},
\end{align}
where $\Pc^{(i)}$ with $i\in\{1,\dots,s\}$ are polyhedral sets with $\interior(\Pc^{(i)})\cap\interior(\Pc^{(j)})=\emptyset$ for all $i\neq j$ that partition the state and input space according to $\Xc \times \Uc=\cup_{i=1}^s\Pc^{(i)}$. Assume that $\xb=\zerob$ is an equilibrium of the system for $\ub=\zerob$. Hence, $\pb^{(i)}=\zerob$ for all $i\in\Ic_0:=\{j\in\{1,\dots,s\} \ | \ \zerob \in \Pc^{(j)}\}$. 
Regarding the sets $\Pc^{(i)}$, we furhter assume that they are given in the hyperplane representation $\Pc^{(i)}:=\{ \xib \in \R^{n+m} \ | \ \Hb^{(i)} \xib \leq \hb^{(i)} \}$ with $\Hb^{(i)}\in\R^{d_i\times (n+m)}$ and $\hb^{(i)}\in\R^{d_i}$ being shorthand notation for $\Hb^{(\Pc^{(i)})}$ and $\hb^{(\Pc^{(i)})}$, respectively. A common approach for the control of PWA systems of the form \eqref{eq:PWA_System} is to solve the optimal control problem (OCP)
\begin{subequations}\label{eq:OCP}
        \begin{align}
        &\hspace{-21mm}V(\xb) \! := \!\!\!\!\!\!\min_{\substack{\hat{\xb}(0),...,\hat{\xb}(N)\\ \hat{\ub}(0),...,\hat{\ub}(N-1)}} 
        \!\!\!\!\!
        \norm{\hat{\xb}(N)}^2_{\Pb} \! + \!\!\! \sum_{\kappa=0}^{N-1} \norm{\hat{\xb}(\kappa)}^2_{\Qb} \! + \! \norm{\hat{\ub}(\kappa)}^2_{\Rb}   \\
        \text{s.t.} \quad \hat{\xb}(0)&=\xb, \\
        \label{eq:PWA_Predict}
        \hat{\xb}(\kappa\!+\!1)\!&=\!\fb_{\text{PWA}}\!(\hat{\xb}(\kappa),\hat{\ub}(\kappa)), \forall \kappa\! \in\! \{0,...,N\!-\!1\!\}, \\
        \label{eq:StateInputCon}
        \left(\hat{\xb}(\kappa),\hat{\ub}(\kappa)\right) & \in \Xc \times \Uc, \ \forall \kappa \in \{0,...,N-1\}, \\
        \label{eq:TerminalCon}
        \hat{\xb}(N) & \in \Tc,
    \end{align}
\end{subequations}
in every time step for the current state $\xb=\xb(k)$. The resulting control law $\fb_{\text{MPC}}:\Fc_{\text{MPC}} \rightarrow \Uc$ is $\fb_{\text{MPC}}(\xb):=\Sb \,\Ub^\ast(\xb)$, where $\Sb:=(\Ib \ \zerob \ \dots \ \zerob) \in \R^{m \times mN}$ serves as a selection matrix and where the set $\Fc_{\text{MPC}}$ contains all $\xb \in \R^n$ for which \eqref{eq:OCP} is feasible.

\subsection{Mixed-integer formulation of PWA system dynamics}
To describe the PWA system dynamics of \eqref{eq:PWA_System} by MI linear constraints, we use techniques from \cite{Bemporad1999} and introduce binary variables $\gammab(k)\in\{0,1\}^s$ that model the relation
\begin{equation}\label{eq:delta1EqXUinRi}
    \gammab_i(k)=1\Leftrightarrow\begin{pmatrix}
        \xb(k) \\
        \ub(k)
    \end{pmatrix} \in \Pc^{(i)}.
\end{equation}
For sets $\Pc^{(i)}$ as above, the MI linear constraints
\begin{subequations}\label{eq:gammaForR}
\begin{align}
    \Hb^{(i)} \begin{pmatrix} \xb(k) \\ \ub(k) \end{pmatrix} &\leq \hb^{(i)} + \oneb M (1-\gammab_i(k)) \\
    \oneb^\top \gammab(k) &= 1, \quad \gammab(k)\in\{0,1\}^s
\end{align}
\end{subequations}
for all $i\in\{1,\dots,s\}$ and all $k\in\{0,\dots,K-1\}$ with $K\in\N$ can be used to reflect \eqref{eq:delta1EqXUinRi}. The constant $M$ is often referred to as big-$M$ and can be chosen according to \cite[Eq.~(20)-(21)]{Bemporad1999}. Now, the binary variable $\gammab(k)$ from \eqref{eq:gammaForR} can be used to describe the PWA system dynamics \eqref{eq:PWA_System} by the MI linear constraints
\begin{subequations}\label{eq:MISys}
\begin{align}
    \nonumber
    \hspace{-4mm}-\oneb M(1\!-\!\gammab_i(k))\!&\leq\!\Ab^{(i)}\xb(k)\!+\!\Bb^{(i)}\ub(k)\!+\!\pb^{(i)}\!-\!\tilde{\xb}^{(i)}(k\!+\!1) \\
    &\leq \oneb M(1-\gammab_i(k)) \\ 
    - \oneb M \gammab_i(k) &\leq \tilde{\xb}^{(i)}(k)\leq \oneb M \gammab_i(k) \\
    \sum\limits_{i=1}^s \tilde{\xb}^{(i)}&(k+1) = \xb(k+1)
\end{align}
\end{subequations}
for all $i\in\{1,\dots,s\}$ and all $k\in\{0,\dots,K-1\}$ (cf. \cite[Sec.~3.1]{Bemporad1999}). Thus, the solution of the MI feasibility problem
\begin{subequations}\label{eq:MIFP_PWASys}
\begin{align}
        &\text{find} \ \Xb_{K+1},\Ub_{K},\tilde{\Xb}_{K}^{(1)},...,\tilde{\Xb}_{K}^{(s)}, \gammab(0),...,\gammab(K\!-\!1)  \\ 
        &\text{s.t.} \quad \text{\eqref{eq:gammaForR} and \eqref{eq:MISys}}, \ \text{with }
\end{align}
\end{subequations}
\begin{equation*}
    \Xb_{K+1}\!\! :=\!\! \begin{pmatrix}\!\!
        \xb(0) \\
        \vdots \\
        \xb(K)\!\!
    \end{pmatrix}\!, 
    \Ub_{K}\!\! :=\!\! \begin{pmatrix}\!\!
        \ub(0) \\
        \vdots \\
        \ub(K-1)\!\!
    \end{pmatrix}\!,
    \tilde{\Xb}_{K}^{(i)}\!\! :=\!\! 
    \begin{pmatrix}\!\!
        \tilde{\xb}^{(i)}(1) \\
        \vdots \\
        \tilde{\xb}^{(i)}(K)\!\!
    \end{pmatrix} 
\end{equation*}
for all $i\in\{1,\dots,s\}$, is such that $\xb(k+1)=\fb_{\text{PWA}}(\xb(k),\ub(k))$ holds for all $k\in\{0,\dots,K-1\}$. By replacing the nonlinear system dynamics \eqref{eq:PWA_Predict} in the OCP with the MI linear constraints \eqref{eq:gammaForR} and \eqref{eq:MISys}, the nonlinear OP \eqref{eq:OCP} becomes a mixed-integer quadratic program (MIQP), which can be solved using standard solvers, e.g., \cite{mosek}. 
However, since the number of binary variables of the resulting MIQP is $sN$, the online solution of \eqref{eq:OCP} in every time step may be intractable for long prediction horizons or PWA systems with a large number of regions. Therefore, we aim for approximations of the control law $\fb_{\text{MPC}}$ that are fast to evaluate and 
ensure stability as well as constraint satisfaction.

\subsection{Mixed-integer formulation of NN with PWA activation}
Neural networks are a common choice for approximating various types of complex control laws, as they can be evaluated fast and trained efficiently on large data sets. In general, a feed-forward-NN with $\ell$ hidden layers and $w_i$ neurons in layer $i$ can be written as a composition of the form 
\begin{equation}\label{eq:NN}
    \Phib(\xb):=\fb^{(\ell+1)}\circ \gb^{(\ell)}\circ \fb^{(\ell)}\circ \dots \circ \gb^{(1)}\circ \fb^{(1)}(\xb).
\end{equation}
Here, the functions $\fb^{(i)}: \R^{w_{i-1}} \rightarrow \R^{p_i w_i}$ for $i \in \{1,\dots,\ell\}$ refer to preactivations, where the parameter $p_i\in \N$ allows to consider ``multi-channel'' preactivations as required for maxout (see \cite{Goodfellow2013}). Moreover, $\gb^{(i)}: \R^{p_i w_i} \rightarrow \R^{w_i}$ stand for activation functions and $\fb^{(\ell+1)}: \R^{w_{\ell}} \rightarrow \R^{w_{\ell+1}}$ reflects  postactivation.
The functions $\fb^{(i)}$ are typically affine, i.e., 
$\fb^{(i)}(\yb^{(i-1)}):=\Wb^{(i)}\yb^{(i-1)}+\bb^{(i)}$, where $\Wb^{(i)}\!\!\in\R^{p_i w_i\times w_{i-1}}\!$ is a weighting matrix, ${\bb^{(i)} \!\!\in \R^{p_i w_i}}$ is a bias vector, and $\yb^{(i-1)}$ denotes the output of the previous layer with $\yb^{(0)}:=\xb \in \R^n$. Various choices for the activation functions have been proposed. However, since we aim for approximations of the PWA control law $\fb_{\text{MPC}}(\xb)$ (see \cite[Thm.~1]{Borrelli2003}), we here focus on the maxout activation function
\begin{equation}
    \label{eq:maxout}
    \gb^{(i)}(\zb^{(i)}):=
    \begin{pmatrix}
        \max \limits_{1\leq j \leq p_i}\big\{\zb_j^{(i)}\big\} \\
        \vdots \\
        \max \limits_{p_i(w_i-1)+1 \leq j \leq p_i w_i}\big\{\zb_{j}^{(i)}\big\}
    \end{pmatrix},
\end{equation}
as the resulting NN have a PWA input-output relation. Moreover, the maxout activation is more general than other PWA activation functions and includes, e.g., ReLU or leaky ReLU, as special cases. We use the shorthand notation $\max\limits_{1\leq j \leq p_i}\big\{\zb_j^{(i)}\big\}:=\max\big\{\zb_1^{(i)},\dots,\zb_{p_i}^{(i)}\big\}$ and we will refer to the resulting NN as a maxout NN. We will assume that the conditions $\Phib(\zerob)=\zerob$ and $\Phib(\xb)\in\Uc$ hold for all $\xb\in\Xc$. Note that these conditions are not restrictive. In fact, for box constraints of the form $\underline{\ub}\leq \ub(k) \leq \overline{\ub}$, they can always be enforced by considering 
\begin{equation*}
    \tilde{\Phib}(\xb)=-\max\{-\max\{\Phib(\xb)-\Phib(\zerob),\underline{\ub}\},-\overline{\ub}\}
\end{equation*}
which is again a maxout NN with two additional layers and weights $\tilde{\Wb}^{(i)}=\Wb^{(i)}$, $\tilde{\bb}^{(i)}=\bb^{(i)}$ for all $i\in\{1,\dots,\ell\}$ and
\begin{align*}
    \tilde{\Wb}^{(\ell+1)}&=\begin{pmatrix}
        \Wb^{(\ell+1)} \\
        \zerob
    \end{pmatrix}, & \tilde{\bb}^{(\ell+1)}&=\begin{pmatrix}
        \bb^{(\ell+1)}-\Phib(\zerob) \\
        \underline{\ub}
    \end{pmatrix}, \\
    \tilde{\Wb}^{(\ell+2)}&=\begin{pmatrix}
        -\Ib_m \\
        \zerob
    \end{pmatrix}, & 
    \tilde{\bb}^{(\ell+2)}&=\begin{pmatrix}
        \zerob \\
        -\overline{\ub}
    \end{pmatrix},
\end{align*}
$\tilde{\Wb}^{(\ell+3)}=-\Ib_m$, and $\tilde{\bb}^{(\ell+3)}=\zerob$. Then, we have $\tilde{\Phib}(\zerob)=\zerob$ and $\tilde{\Phib}(\xb)\in\Uc$ for all $\xb\in\Xc$, as required. Thus, in the following, we focus on constraint satisfaction for the state constraints and assume that the input constraints hold whenever the state constraints hold. A similar method was used in \cite[Prop.~1]{Karg2020Reach} for guaranteed input constraint satisfaction for ReLU NN-based control laws. Methods for input constraint satisfaction for arbitrary polyhedral regions $\Uc$ can be found in \cite[Sec.~III-B]{Markolf2021}. For analyzing a closed-loop system consisting of a PWA system and an NN-based controller, we need an MI formulation of the maxout NN similar to the MI feasibility problem \eqref{eq:MIFP_PWASys}. According to \cite[Lem.~2]{Teichrib2023Ifac} the MI linear constraints
\begin{subequations}\label{eq:MI_Constraints_NN}
\begin{align}
    \qb^{(i)}_l(k)\! &\leq\! \Wb^{(i)}_j \qb^{(i-1)}(k)\! +\! \bb^{(i)}_j\! +\! \overline{b}^{(i)} (1\!-\!\deltab^{(i)}_j(k)), \\
    -\qb^{(i)}_l(k) &\leq -\Wb^{(i)}_j \qb^{(i-1)}(k)\! -\! \bb^{(i)}_j\! -\! \varepsilon(1\!-\!\deltab^{(i)}_j(k)), \\
    \sum\limits_{\jmath\in\Ac^{(i)}_l} &\deltab^{(i)}_\jmath(k) =1, \quad  \\
    \forall j &\in \Ac^{(i)}_l, \ \forall l \in \{1,\dots,w_i\},  \forall i \in \{1,\dots,\ell\},
\end{align} 
\end{subequations}
with $\Ac^{(i)}_l:=\{p_i(l-1)+1,\dots,p_i l\}$ and $\deltab^{(i)}(k)\in\{0,1\}^{p_i w_i}$ are such that the output of the NN is $\Phib(\qb^{(0)}(k))=\Wb^{(\ell+1)}\qb^{(\ell)}(k)+\bb^{(\ell+1)}$. Thus, the desired MI linear constraints are given by \eqref{eq:MI_Constraints_NN}, and the output of a maxout NN can be computed by solving the MI feasibility problem
\begin{subequations}
\begin{align}
    &\text{find} \ \qb^{(1)}(k),\dots,\qb^{(\ell)}(k), \ \text{and} \ \deltab^{(1)}(k),\dots,\deltab^{(\ell)}(k) \\
    &\text{s.t. \quad \eqref{eq:MI_Constraints_NN}} 
\end{align}
\end{subequations}
with $k\in \{0,\dots,K-1\}$.

\section{Closed-loop analysis of PWA systems with neural network-based controller}\label{sec:CLAnalysis}
By combining the MI linear constraints \eqref{eq:gammaForR}, \eqref{eq:MISys} and \eqref{eq:MI_Constraints_NN} for the PWA system dynamics and the maxout NN, respectively, we can describe the closed-loop system consisting of a PWA system and a maxout NN-based controller by the MI feasibility problem
\begin{subequations}\label{eq:MIClosedLoop}
\begin{align}
        \text{find} \ &\Xb_{K+1},\Ub_{K},\tilde{\Xb}_{K}^{(1)},...,\tilde{\Xb}_{K}^{(s)},\gammab(0),...,\gammab(K\!-\!1)\\
        \nonumber
        &\qb^{(1)}(0),\dots,\qb^{(\ell)}(0),\dots,\qb^{(1)}(K-1),\dots,\qb^{(\ell)}(K-1)\\
        \nonumber
        &\deltab^{(1)}(0),\dots,\deltab^{(\ell)}(0),\dots,\deltab^{(1)}(K-1),\dots,\deltab^{(\ell)}(K-1) 
\end{align}
\begin{align}
        \label{eq:MIConClosedLoop1}
        \text{s.t.} \quad \text{\eqref{eq:gammaForR},}& \text{ \eqref{eq:MISys},} \text{ \eqref{eq:MI_Constraints_NN},} \hspace{33mm} \\
        \xb(0)&=\xb \\
        \label{eq:MIConQEqhatX}
        \qb^{(0)}(k)&= \xb(k) \\
        \label{eq:hatUEqWqplusB}
        \ub(k)&=\Wb^{(\ell+1)}\qb^{(\ell)}(k)+\bb^{(\ell+1)}
\end{align}
\end{subequations}
for all $k\in\{0,\dots,K-1\}$, which leads to the following lemma. 
\begin{lem}\label{lem:ClosedLoopMIFP}
    Let $\fb_{\text{PWA}}$ be as in \eqref{eq:PWA_System} and let $\Phib(\xb)$ be a maxout NN as in \eqref{eq:NN}--\eqref{eq:maxout}. Then, any solution to the MI feasibility problem \eqref{eq:MIClosedLoop} is such that 
    \begin{equation}\label{eq:CLSystem}
        \xb(k+1)=\Ab^{(i)}\xb(k)+\Bb^{(i)}\Phib(\xb(k))+\pb^{(i)}
    \end{equation}
    holds for all $k\in\{0,\dots,K-1\}$.
\end{lem}
\begin{proof}
    For $\Xb_{K+1}$ and $\Ub_{K}$ subject to \eqref{eq:gammaForR} and \eqref{eq:MISys} we have $\xb(k+1)=\Ab^{(i)}\xb(k)+\Bb^{(i)}\ub(k)+\pb^{(i)}$ for all $k\in\{1,\dots,K-1\}$, with \eqref{eq:hatUEqWqplusB} this leads to 
    \begin{align*}
        \xb(k\!+\!1)\!=\!\Ab^{(i)}\xb(k)\!+\!\Bb^{(i)}\left(\Wb^{(\ell+1)}\qb^{(\ell)}(k)\!+\!\bb^{(\ell+1)}\right)\!+\!\pb^{(i)}.
    \end{align*}
    Moreover, according to \cite[Lem.~2]{Teichrib2023Ifac} we have $\Phib(\qb^{(0)}(k))=\Wb^{(\ell+1)}\qb^{(\ell)}(k)+\bb^{(\ell+1)}$ for $\qb^{(0)}(k),\dots,\qb^{(\ell)}(k)$ that satisfy \eqref{eq:MI_Constraints_NN}. With \eqref{eq:MIConQEqhatX} this finally gives us $\xb(k+1)=\Ab^{(i)}\xb(k)+\Bb^{(i)}\Phib(\xb(k))+\pb^{(i)}$ for all $k\in\{1,\dots,K-1\}$ and thus proves the claim.
\end{proof}
Lemma~\ref{lem:ClosedLoopMIFP} allows to describe a PWA system with a maxout NN as controller by MI linear constraints. 

\subsection{Reachability analysis}
By adding a cost function to the MI feasibility problem \eqref{eq:MIClosedLoop}, we can analyze the $k$-step reachable set of the closed-loop system by the MILP
\begin{subequations}\label{eq:MILPReachableSets}
\begin{align}
 c_k^\ast(\vb,\Xc):=&\hspace{-5mm}\max_{\substack{\Xb_{K+1},\Ub_{K},\tilde{\Xb}_{K}^{(1)},\dots,\tilde{\Xb}_{K}^{(s)},\gammab(0),\dots,\gammab(K-1)\\ \qb^{(1)}(0),\dots,\qb^{(\ell)}(0),\dots,\qb^{(1)}(K-1),\dots,\qb^{(\ell)}(K-1)\\
 \deltab^{(1)}(0),\dots,\deltab^{(\ell)}(0),\dots,\deltab^{(1)}(K-1),\dots,\deltab^{(\ell)}(K-1)}} \hspace{-5mm}\vb \xb(k) \\
 &\text{s.t.} \quad \eqref{eq:MIConClosedLoop1}, \eqref{eq:MIConQEqhatX}, \eqref{eq:hatUEqWqplusB}, \ \text{and} \,\, \xb(0)\in\Xc
\end{align}
with $\vb\in\R^{1\times n}$.
\end{subequations}
\begin{lem}\label{lem:suppFRk}
    Consider the $k$-step reachable set  
    \begin{align}
        \nonumber
        \Rc_k(\Xc):=\{&\xb(k)\in\R^n \ | \ \xb(k)=\Ab^{(i)}\xb(k\!-\!1)+\\
        \nonumber
        &\Bb^{(i)}\Phib(\xb(k\!-\!1))+\pb^{(i)},\xb(k\!-\!1)\in \Rc_{k-1}(\Xc) \}
    \end{align}
    with $\Rc_0(\Xc):=\Xc$ and let $c_k^\ast(\vb,\Xc)$ be 
    as in \eqref{eq:MILPReachableSets}. 
    Then, the relation $c_k^\ast(\vb,\Xc)=\hb_{\Rc_k(\Xc)}(\vb)$ holds.
\end{lem}
\begin{proof}
    According to Lemma~\ref{lem:ClosedLoopMIFP}, the constraints of the MILP~\eqref{eq:MILPReachableSets} are such that we have $\xb(k+1)=\Ab^{(i)}\xb(k)+\Bb^{(i)}\Phib(\xb(k))+\pb^{(i)}$ for all $k\in\{1,\dots,K-1\}$. Thus, the MILP becomes
    \begin{subequations}\label{eq:MILPReachSetProof}
    \begin{align}
        c_k^\ast(\vb,\Xc)&=\max_{\Xb_{K+1},\Ub_{K}} \vb\xb(k) \\
        \label{eq:MILPReachSetProof_x0}
        \text{s.t.} \quad &\xb(0)\in\Xc=\Rc_0(\Xc) \\
        \label{eq:MILPReachSetProof_x1}
        &\xb(1)=\Ab^{(i)}\xb(0)+\Bb^{(i)}\Phib(\xb(0))+\pb^{(i)}\\
        \label{eq:MILPReachSetProof_x2}
        &\xb(2)=\Ab^{(i)}\xb(1)+\Bb^{(i)}\Phib(\xb(1))+\pb^{(i)}\\
        \nonumber
        &\quad \quad\,\,\,\vdots \\
        \nonumber
        &\xb(k)=\Ab^{(i)}\xb(k-1)+\Bb^{(i)}\Phib(\xb(k-1))+\pb^{(i)}.
    \end{align}
    \end{subequations}
    By definition of $\Rc_1(\Xc)$, we can summarize the constraints \eqref{eq:MILPReachSetProof_x0} and \eqref{eq:MILPReachSetProof_x1} as $\xb(1)\in\Rc_1(\Xc)$. Then, the constraint \eqref{eq:MILPReachSetProof_x2} and $\xb(1)\in\Rc_1(\Xc)$ can be summarized by $\xb(2)\in\Rc_2(\Xc)$. Continuing this way until the $(k+1)$-th constraint, we can summarize all constraints by $\xb(k)\in\Rc_k(\Xc)$. This results in
    \begin{subequations}\label{eq:maxCx}
    \vspace{-4mm}
    \begin{align}
         c_k^\ast(\vb,\Xc)=&\max_{\xb(k)} \vb\xb(k) \\
        \text{s.t.} \quad &\xb(k)\in\Rc_{k}(\Xc), 
    \end{align}
    \end{subequations}
    which is the support function of $\Rc_{k}(\Xc)$ evaluated for the row-vector $\vb$, i.e., $c_k^\ast(\vb,\Xc)=h_{\Rc_{k}(\Xc)}(\vb)$. 
\end{proof}
This result can be used to compute polyhedral over-approximations of the $k$-step reachable set according to the following lemma.
\begin{lem}\label{lem:ReachOverapprox}
    Let $\Rc_k(\Xc)$ be as in Lemma~\ref{lem:suppFRk}.
    Then, the set 
    \begin{equation}\label{eq:overlineRk}
    \overline{\Rc}_k(\Xc):=\{\xb\in\R^n \ | \ \Cb\xb\leq\cb\}
    \end{equation}
    with $\Cb\in\R^{l\times n}$ and $\cb:=\begin{pmatrix} c_k^\ast(\Cb_1,\Xc) & \dots & c_k^\ast(\Cb_l,\Xc) \end{pmatrix}^\top$ is such that $\Rc_k(\Xc)\subseteq\overline{\Rc}_k(\Xc)$ and $h_{\Rc_{k}(\Xc)}(\Cb)=h_{\overline{\Rc}_{k}(\Xc)}(\Cb)$.
\end{lem}
\begin{proof}
    We first prove that $\Rc_k(\Xc)\subseteq\overline{\Rc}_k(\Xc)$ holds, which is the case if and only if $\hb_{\Rc_k(\Xc)}(\Cb)\leq\cb$. According to Lemma~\ref{lem:suppFRk}, we have  $c_k^\ast(\Cb_i,\Xc)=\hb_{\Rc_k(\Xc)}(\Cb_i)$ for all $i \in \{1,\dots,l\}$ and thus $\cb=\hb_{\Rc_k(\Xc)}(\Cb)$ which implies $\hb_{\Rc_k(\Xc)}(\Cb)\leq\cb$. To prove $h_{\Rc_{k}(\Xc)}(\Cb)=h_{\overline{\Rc}_{k}(\Xc)}(\Cb)$, we first note that
    \begin{align*}
        \hb_{\overline{\Rc}_k(\Xc)}(\Cb_i)=\max\limits_{\xb\in\overline{\Rc}_k(\Xc)} \Cb_i \xb&=\max\limits_{\Cb\xb\leq\hb_{\Rc_k(\Xc)}(\Cb)} \Cb_i \xb \\
        &\leq \max\limits_{\Cb_i\xb\leq\hb_{\Rc_k(\Xc)}(\Cb_i)} \Cb_i \xb 
    \end{align*}
    holds and thus $\hb_{\overline{\Rc}_k(\Xc)}(\Cb_i)\leq\hb_{\Rc_k(\Xc)}(\Cb_i)$. Moreover, due to $\Rc_k(\Xc)\subseteq\overline{\Rc}_k(\Xc)$, the inequality $\hb_{\Rc_k(\Xc)}(\Cb_i)\leq\hb_{\overline{\Rc}_k(\Xc)}(\Cb_i)$ holds. In combination, we find $\hb_{\Rc_k(\Xc)}(\Cb_i)=\hb_{\overline{\Rc}_k(\Xc)}(\Cb_i)$ for all $i \in \{1,\dots,l\}$.
\end{proof}
While in general it is computationally demanding or even impossible to compute the exact $k$-step reachable set for a PWA system with an NN-based controller, Lemma~\ref{lem:ReachOverapprox} provides a way to compute over-approximations $\overline{\Rc}_{k}(\Xc)$ of $\Rc_{k}(\Xc)$ which are tight in the direction of the vectors $\Cb_i$. In other words, both sets have the same expansion in the direction of $\Cb_i$, since $h_{\Rc_{k}(\Xc)}(\Cb_i)=h_{\overline{\Rc}_{k}(\Xc)}(\Cb_i)$ holds. 

The following corollary describes a feature 
of the over-approximation $\overline{\Rc}_1(\Xc)$ that is frequently used in this paper. 
\begin{cor}\label{cor:RoverASubseteqRoverB}
    Let $\overline{\Rc}_1(\Xc)$ be defined as in \eqref{eq:overlineRk} and assume that $\Ac\subseteq\Bc$ holds for two polyhedral sets $\Ac\subset\R^n$ and $\Bc\subset\R^n$. Then, $\overline{\Rc}_1(\Ac)\subseteq\overline{\Rc}_1(\Bc)$ holds.  
\end{cor}
\begin{proof}
    By assumption, we have $\Ac\subseteq\Bc\Leftrightarrow\Ac\cap\Bc=\Ac$. In combination with \eqref{eq:MILPReachSetProof} we can conclude that $c^\ast_k(\Cb_i,\Ac)=c^\ast_k(\Cb_i,\Ac\cap\Bc)\leq c^\ast_k(\Cb_i,\Bc)$ holds for all $i\in\{1,\dots,l\}$. The latter is equivalent to $h_{\overline{\Rc}_1(\Ac)}(\Cb)\leq h_{\overline{\Rc}_1(\Bc)}(\Cb) = \hb^{(\overline{\Rc}_1(\Bc))}$ and thus $\overline{\Rc}_1(\Ac)\subseteq\overline{\Rc}_1(\Bc)$ holds.    
\end{proof}

In the remainder of this section, we describe methods that can be used to check whether a given set is positively invariant (PI) for the closed system \eqref{eq:CLSystem} and with which PI sets with various desirable properties, e.g., small or large, can be computed.
\begin{lem}\label{lem:PosInvCheckAndROverOver}
    Let $\overline{\Rc}_1(\Fc)$ be defined as in \eqref{eq:overlineRk} and assume that $\overline{\Rc}_1(\Fc)\subseteq\Fc$ holds for an arbitrary polyhedral set $\Fc\subset\R^n$. Then, 
    \begin{itemize}
        \item[(i)] the set $\Fc$ is PI for the closed-loop system \eqref{eq:CLSystem},
    \end{itemize}
    \begin{itemize}
        \item[(ii)] the set
        \begin{equation}\label{eq:RkOverOver}
            \overline{\Rc_k(\Fc)}:=\underbrace{\overline{\Rc}_1(\overline{\Rc}_1(\dots\overline{\Rc}_1}_{k \ \text{times}}(\Fc)))
        \end{equation}
        is PI with $\Rc_k(\Fc)\subseteq\overline{\Rc_k(\Fc)}\subseteq\Fc$ and $\overline{\Rc}_1(\overline{\Rc_k(\Fc)})\subseteq\overline{\Rc_k(\Fc)}$ for all $k\in\N$.
    \end{itemize}
\end{lem}
\begin{proof}
    (i) Since $\Rc_1(\Fc)\subseteq\overline{\Rc}_1(\Fc)$ holds according to Lemma~\ref{lem:ReachOverapprox}, we have $\Rc_1(\Fc)\subseteq\overline{\Rc}_1(\Fc)\subseteq\Fc $ and consequently $\Rc_1(\Fc)\subseteq\Fc$. By using the definition of $\Rc_1(\Fc)$ we obtain $\Rc_1(\Fc)=\{\xb^+\in\R^n \ | \ \xb^+=\Ab^{(i)}\xb+\Bb^{(i)}\Phib(\xb)+\pb^{(i)},\xb\in \Fc \} \subseteq \Fc$. Thus, for all $\xb \in \Fc$, the successor state $\xb^+$ is in the set $\Fc$, which proves that $\Fc$ is PI for the closed-loop system~\eqref{eq:CLSystem} (cf. \cite[Sec.~3.2]{Blanchini1999}). 
    
    (ii) We prove the claim by induction. For $k=1$ we have $\overline{\Rc_1(\Fc)}=\overline{\Rc}_1(\Fc)$ by definition and thus $\overline{\Rc}_1(\overline{\Rc_1(\Fc)})=\overline{\Rc}_1(\overline{\Rc}_1(\Fc))\subseteq\overline{\Rc}_1(\Fc)=\overline{\Rc_1(\Fc)}\subseteq\Fc$, where we used that $\overline{\Rc_1(\Fc)}=\overline{\Rc}_1(\Fc)\subseteq\Fc$ holds by assumption and thus $\overline{\Rc}_1(\overline{\Rc}_1(\Fc))\subseteq\overline{\Rc}_1(\Fc)$ holds according to Corollary~\ref{cor:RoverASubseteqRoverB}. Consequently, $\overline{\Rc_1(\Fc)}$ is according to (i) PI and $\Rc_1(\Fc)\subseteq\overline{\Rc}_1(\Fc)=\overline{\Rc_1(\Fc)}\subseteq\Fc$ holds. Thus, the induction hypothesis $\overline{\Rc}_1(\overline{\Rc_k(\Fc)}) \subseteq \overline{\Rc_k(\Fc)}$ and $\Rc_k(\Fc)\subseteq\overline{\Rc_k(\Fc)}\subseteq\Fc$ hold for one $k=i$. For $k=i+1$ we have 
    \begin{align*}
        \overline{\Rc}_1(\overline{\Rc_{i+1}(\Fc)})&=\overline{\Rc}_1\left(\overline{\Rc}_1(\overline{\Rc_{i}(\Fc)})\right) \\
        &\subseteq \overline{\Rc}_1(\overline{\Rc_{i}(\Fc)})=\overline{\Rc_{i+1}(\Fc)}.
    \end{align*}
    Moreover, we have $\overline{\Rc_{i+1}(\Fc)}=\overline{\Rc}_1(\overline{\Rc_{i}(\Fc)})\subseteq\overline{\Rc}_1(\Fc)\subseteq\Fc$ and $\Rc_{i+1}(\Fc)=\Rc_{1}(\Rc_{i}(\Fc))\subseteq\overline{\Rc}_{1}(\Rc_i(\Fc))\subseteq\overline{\Rc}_{1}(\overline{\Rc_i(\Fc)})=\overline{\Rc_{i+1}(\Fc)}$ by the induction hypothesis and Corollary~\ref{cor:RoverASubseteqRoverB}. In combination with the proof of (i), we can conclude that $\overline{\Rc_k(\Fc)}$ is a PI set with $\Rc_k(\Fc)\subseteq\overline{\Rc_k(\Fc)}\subseteq\Fc$ and $\overline{\Rc}_1(\overline{\Rc_k(\Fc)})\subseteq\overline{\Rc_k(\Fc)}$ for all $k\in\N$.
\end{proof}
Lemma~\ref{lem:PosInvCheckAndROverOver}, allows computing potentially smaller PI sets $\overline{\Rc_k(\Fc)}$ that are contained in a given larger PI set $\Fc$ and can be reached from all $\xb\in\Fc$ in at most $k$ time steps, since $\Rc_k(\Fc)\subseteq\overline{\Rc_k(\Fc)}$ holds. Note that, for linear systems, there always exits a $k$ such that $\Rc_k(\Fc_{\text{max}})\subseteq\Fc_{\text{min}}$ holds if $\Fc_{\text{max}}$ and $\Fc_{\text{min}}$ are PI sets with $\Fc_{\text{min}}\subseteq\Fc_{\text{max}}$. Thus, $\Fc_{\text{min}}$ is reachable for all $\xb \in \Fc_{\text{max}}$. However, for PWA systems, there may exist PI sets with $\Fc_{\text{min}}\subseteq\Fc_{\text{max}}$ for which $\Fc_{\text{min}}$ is not reachable from all $\xb\in\Fc_{\text{max}}$, i.e., $\Rc_k(\Fc_{\text{max}})\not\subseteq\Fc_{\text{min}}$ for all $k$. Therefore, $\Rc_k(\Fc)\subseteq\overline{\Rc_k(\Fc)}$ in Lemma~\ref{lem:PosInvCheckAndROverOver} (ii) is crucial and does not follow from the fact that $\overline{\Rc_k(\Fc)}$ and $\Fc$ are PI sets with $\overline{\Rc_k(\Fc)}\subseteq\Fc$. In addition, using $\overline{\Rc_k(\Fc)}$ has several computational advantages over $\overline{\Rc}_k(\Fc)$. The set $\overline{\Rc_k(\Fc)}$ is guaranteed to be PI if $\Fc$ is PI. While $\overline{\Rc}_k(\Fc)$ may not be PI for some $k$ even if $\Fc$ is PI. More importantly, $\overline{\Rc_k(\Fc)}$ relies on repeatedly computing over-approximations of one-step reachable sets. This is computationally more efficient than computing the over-approximation of the $k$-step reachable set $\overline{\Rc}_k(\Fc)$.

\subsection{Convergence to a set containing the origin}\label{sec:ConvToFmin}
We first aim for the computation of a large PI set $\Fc_{\text{max}}$ with $\Fc_{\text{max}}\subseteq\Xc$ and thus $\Phib(\xb)\subseteq\Uc$ for all $\xb\in\Fc_{\text{max}}$, which will serve as feasible set for the NN-based controller. Ideally, $\Fc_{\text{max}}$ should be the maximal PI set, which is the union of all sets $\Fc$ satisfying $\Rc_1(\Fc)\subseteq\Fc$. However, an exact computation of the maximal PI set for \eqref{eq:CLSystem} requires knowledge of all affine segments and polyhedral sets of the PWA function represented by the NN \cite[Alg.~4.1]{Rakovic2004}. Computing these affine segments and sets is computationally intractable even for small NN. Therefore, the following algorithm provides a method for computing a polyhedral inner approximation of the maximum PI set.       

\begin{alg} Compute a large PI set $\Fc_{\text{max}}\subseteq\Xc$.
\label{alg:Fmax}
\begin{algorithmic}[1]
\State initialize $\Fc \gets \Xc$ and $\Cb \gets \Hb^{(\Xc)}$
\While{$\overline{\Rc}_1(\Fc) \not\subseteq\Fc$}
\State  set $\Fc \gets \overline{\Rc}_1(\Fc) \cap \Xc$
\EndWhile
\State set $\Fc_{\text{max}}\gets\Fc$
\State \Return $\Fc_{\text{max}}$
\end{algorithmic}
\end{alg}
In contrast to \cite[Alg.~4.1]{Rakovic2004}, we use here in line 3 of the algorithm the over-approximation \eqref{eq:RkOverOver} of the one-step reachable set, which allows the application to PWA systems of the form \eqref{eq:CLSystem}. 
\begin{prop}\label{prop:Fmax}
    Let $\overline{\Rc_k(\Xc)}$ be defined as in \eqref{eq:RkOverOver} and assume that there exists a $\overline{k}$ such that $\overline{\Rc_{\overline{k}+1}(\Xc)}\subseteq\overline{\Rc_{\overline{k}}(\Xc)}\subseteq\Xc$ holds. Then, Algorithm~\ref{alg:Fmax} terminates with a PI set $\Fc_{\text{max}}\subseteq\Xc$ after at most $\overline{k}$ iterations.
\end{prop}
\begin{proof}
    The set computed in the $(k+1)$-th iteration of Algorithm~\ref{alg:Fmax} is $\Fc_{k+1}=\overline{\Rc}_1(\Fc_k)\cap\Xc \ \text{with} \ \Fc_0=\Xc$. We first note, that by construction $\Fc_{k+1}\subseteq\Fc_{k}\subseteq\Xc$ and $\Fc_{k}\subseteq\overline{\Rc_{k}(\Xc)}$ hold for all $k\in\N$. Thus, we have $\overline{\Rc}_1(\Fc_{\overline{k}})\subseteq\overline{\Rc}_1(\overline{\Rc_{\overline{k}}(\Xc)})=\overline{\Rc_{\overline{k}+1}(\Xc)}\subseteq\overline{\Rc_{\overline{k}}(\Xc)}\subseteq\Xc$, which implies $\overline{\Rc}_1(\Fc_{\overline{k}})=\overline{\Rc}_1(\Fc_{\overline{k}})\cap\Xc=\Fc_{\overline{k}+1}$. Consequently, $\overline{\Rc}_1(\Fc_{\overline{k}})=\Fc_{\overline{k}+1}\subseteq\Fc_{\overline{k}}\subseteq\Xc$ holds, which proves that $\Fc_{\overline{k}}$ is a PI set contained in $\Xc$. Thus Algorithm~\ref{alg:Fmax} will terminate after at most $\overline{k}$ iterations, since $\overline{\Rc}_1(\Fc_{\overline{k}})\subseteq\Fc_{\overline{k}}$ holds.
\end{proof}
\begin{rem}\label{rem:Fmax}
The condition from Proposition~\ref{prop:Fmax}, which guarantees finite termination of Algorithm~\ref{alg:Fmax}, requires that a PI and reachable set $\overline{\Rc_{\overline{k}}(\Xc)}$ exists that is contained in $\Xc$. Surely, the set $\overline{\Rc_{\overline{k}}(\Xc)}$ could also be used as feasible set for the NN-based controller. However, since $\overline{k}$ is an upper bound, the Algorithm~\ref{alg:Fmax} will typically terminate in less than $\overline{k}$ iterations with a PI set $\Fc_{\text{max}}$ that is larger than $\overline{\Rc_{\overline{k}}(\Xc)}$.     
\end{rem}
With Proposition~\ref{prop:Fmax}, we can guarantee constraint satisfaction of the closed-loop system with NN-based controller \eqref{eq:CLSystem} for all trajectories starting from $\Fc_{\text{max}}$, since $\Fc_{\text{max}}$ is a PI set contained in $\Xc$. Moreover, by assumption, we then further have $\Phib(\xb)\in\Uc$. To show that the closed-loop system with NN-based controller reaches and remains in a possibly small set containing the origin for all $\xb(0)\in\Fc_{\text{max}}$, we will next state a result that allows the computation of small PI sets.
\begin{prop}\label{prop:Fmin}
    Let $\overline{\Rc_k(\Fc_{\text{max}})}$ be defined as in \eqref{eq:RkOverOver} with $\Fc_{\text{max}}$ being the PI set computed with Algorithm~\ref{alg:Fmax} and assume that there exists a $k^\ast$ such that 
    \begin{equation}\label{eq:epsCondFmax}
        \frac{1}{1+\epsilon}\overline{\Rc_{k^\ast}(\Fc_{\text{max}})}\subseteq\overline{\Rc}_1\left(\frac{1}{1+\epsilon}\overline{\Rc_{k^\ast}(\Fc_{\text{max}})}\right)
    \end{equation} 
    holds for a given $\epsilon>0$. Then, the set
    \begin{equation}\label{eq:Fmin}
        \Fc_{\text{min}}:=\overline{\Rc_{k^\ast}(\Fc_{\text{max}})}
    \end{equation}
    is PI with 
    \begin{equation}\label{eq:epsInclusion}
        \overline{\Rc_{\infty}(\Fc_{\text{max}})}\subseteq\overline{\Rc_{k^\ast}(\Fc_{\text{max}})}\subseteq(1+\epsilon)\overline{\Rc_{\infty}(\Fc_{\text{max}})},
    \end{equation}
    where $\overline{\Rc_{\infty}(\Fc_{\text{max}})}$ is short for $\lim_{k\to\infty}\overline{\Rc_{k}(\Fc_{\text{max}})}$.    
\end{prop}
\begin{proof}
    Since $\overline{\Rc}_1(\Fc_{\text{max}})\subseteq\Fc_{\text{max}}$ holds for $\Fc_\text{max}$ computed with Algorithm~\ref{alg:Fmax}, the set $\Fc_{\text{min}}=\overline{\Rc_{k^\ast}(\Fc_{\text{max}})}$ is PI for all $k^\ast\in\N$ according Lemma~\ref{lem:PosInvCheckAndROverOver} (ii). It remains to prove that the inclusion \eqref{eq:epsInclusion} holds. The left-hand side of the inclusion holds since we have $\overline{\Rc_{k+1}(\Fc_{\text{max}})}=\overline{\Rc}_1(\overline{\Rc_{k}(\Fc_{\text{max}})})\subseteq\overline{\Rc_{k}(\Fc_{\text{max}})}$ for all $k\in\N$ as apparent from Lemma~\ref{lem:PosInvCheckAndROverOver} (ii) and thus $\overline{\Rc_{\infty}(\Fc_{\text{max}})}\subseteq\overline{\Rc_{k^\ast}(\Fc_{\text{max}})}$. For the right-hand side, we prove that $\overline{\Rc_{k^\ast}(\Fc_{\text{max}})}\subseteq(1+\epsilon)\overline{\Rc_{k^\ast+k}(\Fc_{\text{max}})}$ holds for all $k\in\N_0$. For $k=0$ we have $\overline{\Rc_{k^\ast}(\Fc_{\text{max}})}\subseteq(1+\epsilon)\overline{\Rc_{k^\ast}(\Fc_{\text{max}})}$. Thus, $\overline{\Rc_{k^\ast}(\Fc_{\text{max}})}\subseteq(1+\epsilon)\overline{\Rc_{k^\ast+i}(\Fc_{\text{max}})}$ holds for one $i\in\N_0$. For $k=i+1$ we obtain
    \begin{align*}
        \overline{\Rc_{k^\ast}(\Fc_{\text{max}})}&\subseteq(1+\epsilon)\overline{\Rc}_1\left(\frac{1}{1+\epsilon}\overline{\Rc_{k^\ast}(\Fc_{\text{max}})}\right)\\
        &\subseteq(1+\epsilon)\overline{\Rc}_1(\overline{\Rc_{k^\ast+i}(\Fc_{\text{max}})})\\
        &=(1+\epsilon)\overline{\Rc_{k^\ast+i+1}(\Fc_{\text{max}})}
    \end{align*}
    by using assumption \eqref{eq:epsCondFmax}, the induction hypothesis, and Corollary~\ref{cor:RoverASubseteqRoverB}. Which proves that $\overline{\Rc_{k^\ast}(\Fc_{\text{max}})}\subseteq(1+\epsilon)\overline{\Rc_{k^\ast+k}(\Fc_{\text{max}})}$ holds for all $k\in\N_0$. In summary, we have $\overline{\Rc_{\infty}(\Fc_{\text{max}})}\subseteq\overline{\Rc_{k^\ast}(\Fc_{\text{max}})}\subseteq(1+\epsilon)\lim\limits_{k\to\infty}\overline{\Rc_{k^\ast+k}(\Fc_{\text{max}})}$, which completes the proof.
\end{proof}
\begin{rem}
Condition \eqref{eq:epsCondFmax} has a rather intuitive interpretation. According Lemma~\ref{lem:PosInvCheckAndROverOver} (ii) we have $\overline{\Rc}_1(\overline{\Rc_k(\Fc_{\text{max}})})\subseteq\overline{\Rc_k(\Fc_{\text{max}})}$ for all $k\in\N$ and thus the reachable sets $\overline{\Rc_k(\Fc_{\text{max}})}$ usually becomes smaller when $k$ grows and finally converge to $\overline{\Rc_{\infty}(\Fc_{\text{max}})}$. However, if the reachable sets starting from $\Fc$ grow, i.e., if we have $\Fc\subseteq\overline{\Rc}_1(\Fc)$, then this implies $\Fc\subseteq\overline{\Rc_{\infty}(\Fc_{\text{max}})}$. Thus condition \eqref{eq:epsCondFmax} states that $k^\ast$ is such that a small scaling of the set $\overline{\Rc_{k^\ast}(\Fc_{\text{max}})}$ with $1/(1+\epsilon)$ leads to a growing reachable set (cf. right-hand side of \eqref{eq:epsCondFmax}) and thus $1/(1+\epsilon)\overline{\Rc_{k^\ast}(\Fc_{\text{max}})}\subseteq\overline{\Rc_{\infty}(\Fc_{\text{max}})}$, which is exactly what we aim for. For the limit case $\epsilon\to0$ condition \eqref{eq:epsCondFmax} becomes $\overline{\Rc_k(\Fc_{\text{max}})}\subseteq\overline{\Rc}_1(\overline{\Rc_k(\Fc_{\text{max}})})$ which is equivalent to $\overline{\Rc_k(\Fc_{\text{max}})}=\overline{\Rc_{k+1}(\Fc_{\text{max}})}=\overline{\Rc_{\infty}(\Fc_{\text{max}})}$. This means that for $\epsilon\to 0$ we need finite determinedness of $\overline{\Rc_{\infty}(\Fc_{\text{max}})}$ or $k^\ast\to\infty$ for \eqref{eq:epsCondFmax} to hold. For the other limit case $\epsilon\to\infty$, the left-hand side of \eqref{eq:epsCondFmax} becomes a set only containing the origin. Moreover, since $\Phib(\zerob)=\zerob$ and $\fb_{\text{PWA}}(\zerob,\zerob)=\zerob$ holds, the right-hand side of \eqref{eq:epsCondFmax} also becomes a set only containing the origin and thus \eqref{eq:epsCondFmax} holds for all $k^\ast\geq0$. This means that with $\epsilon$, we can influence the number of computation steps needed to compute \eqref{eq:Fmin}.   
\end{rem}
We now combine the previous results to state the main contribution of this section. 

\begin{thm}\label{thm:BoundedStab}
    Let $\Fc_{\text{max}}$ be the set computed with Algorithm~\ref{alg:Fmax} and $\Fc_{\text{min}}$ by defined as in Proposition \ref{prop:Fmin}. Then, the system \eqref{eq:PWA_System} with $\ub(k)=\Phib(\xb(k))$ is ultimately bounded in $\Fc_{\text{min}}$, uniformly in $\Fc_{\text{max}}$. Moreover, $\xb(k)\in\Xc$ and $\ub(k)=\Phib(\xb(k))\in\Uc$ holds for all $k\in \N_0$.
\end{thm}
\begin{proof}
    Since $\Fc_{\text{max}}$ is a PI set for the closed-loop system \eqref{eq:CLSystem}, we have for all $\xb(0)\in\Fc_{\text{max}}$ that $\xb(k)\in\Fc_{\text{max}}\subseteq\Xc$ holds for all $k\in\N$. Moreover, by assumption we have $\Phib(\xb)\in\Uc$ for all $\xb\in\Xc$ and thus $\Phib(\xb(k))\in\Uc$ for all $k\in\N_0$. According to Lemma~\ref{lem:PosInvCheckAndROverOver} (ii) and Proposition~\ref{prop:Fmin}, $\Fc_{\text{min}}$ is a PI set with $\Rc_{k^\ast}(\Fc_{\text{max}})\subseteq\overline{\Rc_{k^\ast}(\Fc_{\text{max}})}=\Fc_{\text{min}}$. Thus, $\xb(k)\in\Rc_{k^\ast}(\Fc_{\text{max}})\subseteq\Fc_{\text{min}}$ holds for all $k\geq k^\ast$ and for all $\xb(0)\in\Fc_{\text{max}}$. This proves, that the system \eqref{eq:PWA_System} with $\ub(k)=\Phib(\xb(k))$ is ultimately bounded in $\Fc_{\text{min}}$, uniformly in $\Fc_{\text{max}}$.
\end{proof}

\subsection{Stability of the origin}\label{sec:CLAnalysis.Origin}
Consider the modified dual-mode control law
\begin{equation}\label{eq:SwitchingPi}
\pib(\xb) := \left\{
\begin{array}{cl}
 \kappab(\xb) & \text{if} \ \xb\in s\Fc_0, \\
\Phib(\xb) & \text{if} \ \xb\in\Fc_{\text{max}}\setminus s\Fc_0,
\end{array}
\right. 
\end{equation} 
where $\Fc_{\text{max}}$ is the PI set computed with Algorithm~\ref{alg:Fmax} and $\kappab(\xb)$ is a stabilizing PWA control law with region of attraction $\Fc_0:=\{\xb\in\R^n \ | \ \xb^\top\Sb\xb \leq \xi^\ast \}$ for the PWA system \eqref{eq:PWA_System}. The set $\Fc_0$ is such that $\xi^\ast$ is the largest $\xi$ for which $\{\xb\in\R^n \ | \ \xb^\top\Sb\xb \leq \xi,\xi>0 \} \subseteq \cup_{i\in\Ic_0}\Pc^{(i)}$ holds. In addition, $0<s\leq 1$ is a scaling factor that is chosen such that $\Fc_{\text{min}}\subseteq s\Fc_0$ holds. The control law $\kappab(\xb)$, $\Sb$, and $\Fc_0$ can be computed by, e.g., solving the linear matrix inequality (LMI) described in \cite[Eq.~(9)-(10)]{Mignone2000}. Then $V(\xb):=\xb^\top\Sb\xb$ is a common Lyapuov function for the system \eqref{eq:PWA_System} with $\ub=\kappab(\xb)$ for all $\xb\in\cup_{i\in\Ic_0}\Pc^{(i)}$. Thus the set $s \Fc_0$ is PI (see \cite[Sec.~3.2]{Mignone2000}).
\begin{thm}\label{thm:AsymStab}
    Let $\pib(\xb)$ and $\Fc_{\text{min}}$ be defined as in \eqref{eq:SwitchingPi} and Proposition~\ref{prop:Fmin}, respectively and assume that $\Fc_{\text{min}}\subseteq s\Fc_0$ with $0<s\leq 1$ holds. Then, the origin is asymptotically stable for the system \eqref{eq:PWA_System} with $\ub(k)=\pib(\xb(k))$. The region of attraction is $\Fc_{\text{max}}$.
\end{thm}
\begin{proof}
    We distinguish the two cases $\xb\in s\Fc_0$ and $\xb\in\Fc_{\text{max}}\setminus s\Fc_0$. In the first case, we have $\ub=\pib(\xb)=\kappab(\xb)$, which asymptotically stabilizes the origin of the PWA system while ensuring $\xb(k)\in s\Fc_0\subseteq\Xc$ for all $k\in\N_0$ (\cite[Lem.~1]{Mignone2000}). To prove asymptotic stability for states $\xb\in\Fc_{\text{max}}\setminus s\Fc_0$ it remains to show that trajectories starting in $\xb\in\Fc_{\text{max}}\setminus s\Fc_0$ enter $s\Fc_0$ after a finite number of time-steps by applying $\ub=\pib(\xb)=\Phib(\xb)$. By assumption we have $\Rc_{k^\ast}(\Fc_{\text{max}})\subseteq\overline{\Rc_{k^\ast}(\Fc_{\text{max}})}=\Fc_{\text{min}}\subseteq s\Fc_0$. Thus, for all $\xb(0)\in\Fc_{\text{max}}\subseteq\Xc$ there exist a $\tilde{k}\leq k^\ast$ such that $\xb(\tilde{k})\in s\Fc_0$ holds when applying $\ub(k)=\Phib(\xb(k))$ for all $k<\tilde{k}$. Since $s\Fc_0$ is PI for \eqref{eq:PWA_System} with $\ub=\kappab(\xb)$ we have $\xb(k)\in s\Fc_0$ and consequently $\ub(k)=\kappab(\xb(k))$ for all $k\geq\tilde{k}$, which asymptotically stabilizes the system. In summary, the control law \eqref{eq:SwitchingPi} asymptotically stabilizes the system \eqref{eq:PWA_System} for all $\xb\in\Fc_{\text{max}}$.
\end{proof}
The parameter $s$ is a design parameter which can be used to scale the set in which the controller $\kappab(\xb)$ acts. Typically, we want to have $\pib(\xb)=\Phib(\xb)$ as long as possible and thus choose $s=\min\limits_{\Fc_{\text{min}}\subseteq c\Fc_0} c$, which is the smallest $s$ that satisfies the conditions of Theorem~\ref{thm:AsymStab}.

\begin{figure*}[ht!]
\centering
\subfigure[\hspace{0mm}]{\includegraphics[trim={0cm 0cm 0cm 0cm},clip,scale=0.51]{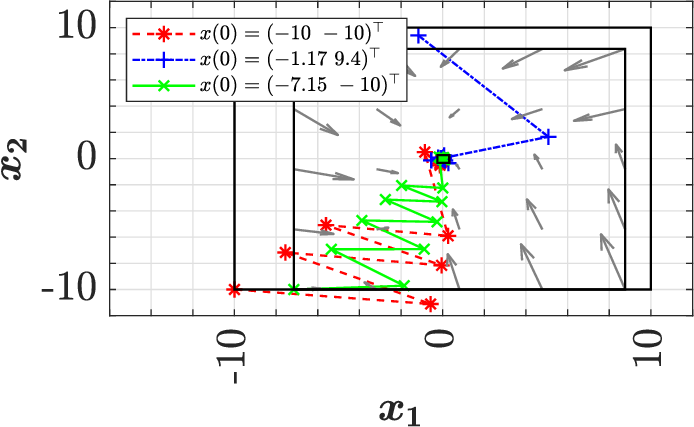}}
\subfigure[\hspace{0mm}]{\includegraphics[trim={0.7cm 0cm 0cm 0cm},clip,scale=0.51]{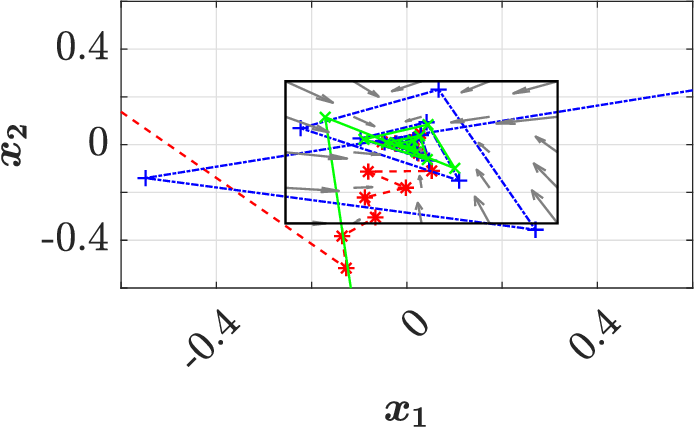}}
\subfigure[\hspace{0mm}]{\includegraphics[trim={0.7cm 0cm 0cm 0cm},clip,scale=0.51]{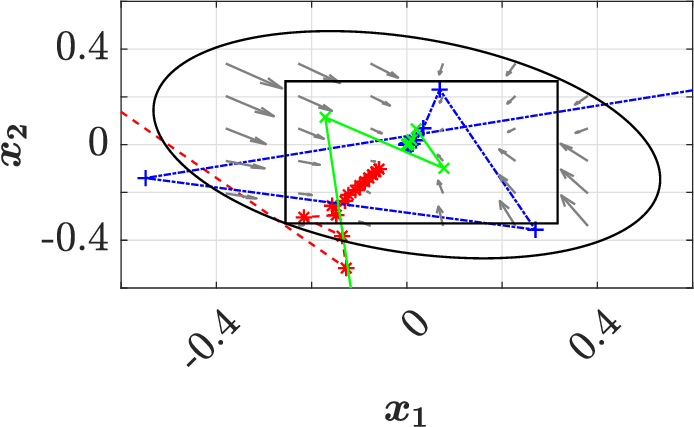}}
\caption{Trajectories of the closed-loop system with the NN-based controller and the sets $\Xc$, $\Fc_{\text{max}}$, and $\Fc_{\text{min}}$, respectively. The initial states of the trajectories are specified in the legend and apply to all figures.}
\label{fig:trajectories}
\end{figure*}

\section{Case study}\label{sec:Example}
Consider the OCP \eqref{eq:OCP} with $\Qb=\Pb=\Ib$, $\Rb=1$ and $N=10$ for a PWA system of the form \eqref{eq:PWA_System} with 
\begin{align*}
    \Ab^{(1)}&=\begin{pmatrix}
        -0.04 & -0.461 \\
        -0.139 & 0.341
    \end{pmatrix}, 
    \Ab^{(2)}=\begin{pmatrix}
        0.936 & 0.323 \\
        0.788 & -0.049
    \end{pmatrix}, \\
    \Ab^{(3)}&=\begin{pmatrix}
        -0.857 & 0.815 \\
        0.491 & 0.62
    \end{pmatrix},
        \Ab^{(4)}=\begin{pmatrix}
        -0.022 & 0.644 \\
        0.758 & 0.271
    \end{pmatrix},\\
    \Bb^{(1)}&=\Bb^{(2)}=\Bb^{(3)}=\Bb^{(4)}=\begin{pmatrix}
        1 & 0
    \end{pmatrix}^\top,
\end{align*}
and $\Hb^{(1)}=\begin{pmatrix}
        -1 & 0 \\ 0 & -1
    \end{pmatrix},
    \Hb^{(2)}=\begin{pmatrix}
        -1 & 0 \\ 0 & 1
    \end{pmatrix}$, \\
$\Hb^{(3)}=-\Hb^{(1)},\Hb^{(4)}=-\Hb^{(2)},$ as well as $\hb^{(1)}=\hb^{(2)}=\hb^{(3)}=\hb^{(4)}=\zerob$ from \cite[Sec.~7]{Mignone2000} with the additional constraints $\norm{\xb(k)}_\infty\leq 10$ for all $k\in\{0,\dots,N\}$ and $|u(k)|\leq 1$ for all $k\in\{0,\dots,N-1\}$. We solved the OCP for $1000$ randomly sampled $\xb\in\Xc$ to generate a data set with the feasible points $(\xb^{(i)},\fb_{\text{MPC}}(\xb^{(i)}))$. This data set is used to train a $3\times 3$ maxout NN, i.e., $\ell=3$ and $w_i=3$ for all $i\in\{1,2,3\}$.
Figure~\ref{fig:trajectories} illustrates trajectories of the closed-loop system with the resulting NN-based controller and different PI sets. In (a), the outer box $[-10,10]\times[-10,10]$ represents the state constraints $\Xc$. The next smaller box $[-7.15,8.76]\times[-10,8.37]$ is $\Fc_{\text{max}}$ in which the gray arrows in (a) and (b) represent the evolution of the closed-loop system with NN-based controller, i.e., $\ub=\Phib(\xb)$. The set $\Fc_{\text{max}}$ is computed according to Algorithm~\ref{alg:Fmax}. The computation terminates after $1$ iteration, whereas we have $\overline{k}=5\geq1$, which is in line with Remark~\ref{rem:Fmax}. Figure~\ref{fig:trajectories}~(b) shows a detailed view of a region around the origin of (a). The smallest box $[-0.25,0.32]\times[-0.33,0.27]$ around the origin is $\Fc_{\text{min}}$. This set is computed according to Proposition~\ref{prop:Fmin} with $\epsilon=10^{-3}$, which results in $k^\ast=81$. Therefore, we have $\Fc_{\text{min}}=\overline{\Rc_{81}(\Fc_{\text{max}})}$ with $\overline{\Rc_{\infty}(\Fc_{\text{max}})}\subseteq\overline{\Rc_{81}(\Fc_{\text{max}})}\subseteq (1+10^{-3}) \overline{\Rc_{\infty}(\Fc_{\text{max}})}$. The Figures~\ref{fig:trajectories} (a) and (b) confirm Theorem~\ref{thm:BoundedStab}, since it can be observed that the trajectory starting in $\Fc_{\text{max}}$, marked with green x, remains in that set and enters $\Fc_{\text{min}}$ after a finite number of time steps (cf. (b)). For the trajectory, marked with red asterisk, we have $\xb(0)\not\in\Fc_{\text{max}}$, which leads to a constraint violation of $\xb(1)$. However, since $\Fc_{\text{max}}$ is only an inner approximation of the maximum PI set contained in $\Xc$, there also exist trajectories (blue plus) with $\xb(0)\not\in\Fc_{\text{max}}$, that enter $\Fc_{\text{min}}$ while satisfying state and input constraints. The results of Section~\ref{sec:CLAnalysis.Origin} are illustrated in Figure~\ref{fig:trajectories} (c). There three trajectories with the same $\xb(0)$ as in (a) are shown for the system \eqref{eq:PWA_System} with $\ub=\pib(\xb)$. The ellipse is the region $s\Fc_0$ with $s=5.32\times 10^{-2}$, in which, according to \eqref{eq:SwitchingPi}, the control law $\ub=\kappab(\xb)$ is applied and asymptotically stabilizes the system. The gray arrows represent the evolution of the corresponding closed-loop system. For the green trajectory, which starts in $\Fc_{\text{max}}$, we can guarantee that it enters $s\Fc_0$ after at most $k^\ast$ time steps and converges to the origin.

\section{Conclusions}\label{sec:Conclusions}
We presented various methods for analyzing PWA systems with an NN-based controller. The combination of the methods leads to the main results of the paper, Theorems~\ref{thm:BoundedStab} and \ref{thm:AsymStab}, which allow certifying stability and constraint satisfaction based on PI sets. 
The proposed approaches all build on the computation of over-approximations of $k$-step reachable sets. This fact indicates a possible direction for future research. In fact, it may be feasible 
to replace the MILP~\eqref{eq:MILPReachableSets} with another OP without binary variables that can be solved faster at the cost of a more conservative over-approximation. Such a modification would not affect the validity of the results since the results of Section~\ref{sec:CLAnalysis} hold independent of the method used to compute the over-approximations.


\end{document}